\documentclass{pasj01}
\input{pasjsup.sty}

\begin{document}

\author{Mariko \textsc{Kimura}, \altaffilmark{1,*}
        Taichi \textsc{Kato}, \altaffilmark{1}
        Akira \textsc{Imada}, \altaffilmark{2}
        Kai \textsc{Ikuta}, \altaffilmark{1}
        Keisuke \textsc{Isogai}, \altaffilmark{1}
        Pavol A. \textsc{Dubovsky}, \altaffilmark{3}
        Seiichiro \textsc{Kiyota}, \altaffilmark{4}
        Roger D. \textsc{Pickard}, \altaffilmark{5}
        Ian \textsc{Miller}, \altaffilmark{6}
        Elena P. \textsc{Pavlenko}, \altaffilmark{7}
        Aleksei A. \textsc{Sosnovskij}, \altaffilmark{7}
        Shawn \textsc{Dvorak}, \altaffilmark{8}
        Daisaku \textsc{Nogami} \altaffilmark{1}
        }
\email{mkimura@kusastro.kyoto-u.ac.jp}

\altaffiltext{1}{Department of Astronomy, Graduate School 
of Science, Kyoto University, Oiwakecho, Kitashirakawa, 
Sakyo-ku, Kyoto 606-8502}

\altaffiltext{2}{Kwasan and Hida Observatories, Kyoto 
University, Yamashina, Kyoto 607-8471}

\altaffiltext{3}{Vihorlat Observatory, Mierova 4, Humenne, 
Slovakia}

\altaffiltext{4}{Variable Star Observers League in Japan 
(VSOLJ), 7-1 Kitahatsutomi, Kamagaya, Chiba 273-0126}

\altaffiltext{5}{The British Astronomical Association, 
Variable Star Section (BAA VSS), Burlington House, 
Piccadilly, London, W1J 0DU, UK}

\altaffiltext{6}{Furzehill House, Ilston, Swansea, 
SA2 7LE, UK}

\altaffiltext{7}{Crimean Astrophysical Observatory, 
298409, Nauchny, Republic of Crimea}

\altaffiltext{8}{Rolling Hills Observatory, 1643 
Nightfall Drive, Clermont, Florida 34711, USA}

\title{Unexpected Superoutburst and Rebrightening of 
AL Comae Berenices in 2015}

\Received{} \Accepted{}


\SetRunningHead{Kimura et al.}{AL Com in 2015}

\maketitle

\begin{abstract}
   In 2015 March, the notable WZ Sge-type dwarf nova AL Com 
exhibited an unusual outburst with a recurrence time of 
${\sim}$1.5 yr, which is the shortest interval of superoutbursts 
among WZ Sge-type dwarf novae. 
Early superhumps 
in the superoutburst light curve were absent, and 
a precursor was observed at the onset of the superoutburst 
for the first time in WZ Sge-type dwarf novae. 
The present superoutburst can be interpreted 
as a result of the condition that the disk radius 
barely reached the 3:1 resonance radius, but 
did not reach the 2:1 resonance one. 
Ordinary superhumps immediately grew following the precursor.  
The initial part of the outburst is indistinguishable from those 
of superoutbursts of ordinary SU UMa-type dwarf novae.
This observation supports the interpretation that the 
2:1 resonance suppresses a growth of ordinary superhumps. 
The estimated superhump period and superhump period derivative 
are $P_{\rm sh}$ = 0.0573185(11) d and $P_{\rm dot} = +1.5(3.1) 
\times 10^{-5}$, respectively. These values indicate 
that the evolution of ordinary superhumps is the same as 
\textcolor{black}{that in} past superoutbursts with much larger 
extent. Although the light curve during the plateau stage was 
typical for an SU UMa-type dwarf nova, this superoutburst showed 
a rebrightening, together with a regrowth of the superhumps. 
The overall light curve of the rebrightening was the almost the 
same as those observed in previous rebrightenings. 
This implies that the rebrightening type is inherent in the 
system.
\end{abstract}

\section{Introduction}

   WZ Sge-type stars are an extreme subclass of dwarf novae 
\citep{osa95wzsge}. The origin of unique light variations, as 
well as their evolutionary status, are still in debate
(e.g., \cite{osa03DNoutburst,nak14j0754j2304}; see 
\cite{kat15wzsge} for a review). Their observational 
properties are that (1) the amplitude of the superoutburst is large, 
exceeding ${\sim}$ 6 mag \citep{how95TOAD}, (2) the interval 
between the successive superoutbursts (supercycle) is unusually 
long, typically with a time scale of decades \citep{kat01hvvir}, 
(3) modulations with double-peaked profiles 
called early superhumps are observed in the early phase of the 
superoutburst \citep{kat02wzsgeESH}, and (4) after the end of the 
plateau stage, single or multiple rebrightenings are observed 
\citep{ima06tss0222}.  Among them, properties 
(3) and (4) are considered to be unique to WZ Sge-type dwarf novae 
and to be the defining characteristics (see \cite{kat15wzsge}).

   In recent years, our understanding on SU UMa-type 
and WZ Sge-type dwarf novae has been improved by using extensive 
photometric data obtained by ground-based and space telescopes 
\textcolor{black}{\citep{sti10v344lyr,woo11v344lyr,osa13kepler,osa14kepler}}.
\citet{Pdot} studied superhump period changes and concluded that 
the $O - C$ diagram of superhump maximum timings is composed of 
three stages: stage A in which $P_{\rm sh}$ (the superhump period) 
is long and constant, stage B in which $P_{\rm sh}$ changes as the 
superoutburst proceeds, and stage C in which $P_{\rm sh}$ is short 
and constant. Also, \citet{Pdot5} classified rebrightenings 
into five types according to the profiles of the light curves. The 
rebrightening type includes type-A (long duration rebrightening), 
type-B (multiple rebrightenings), type-C (single rebrightening), 
type-D (no rebrightening), and type-E (double superoutbursts). 
Most recently, \citet{kat15wzsge} published a comprehensive review 
of WZ Sge-type dwarf novae, in which he suggested that period 
derivative of superhumps and rebrightening type reflect the 
evolutional sequence of the system.

   In the case of WZ Sge-type stars, no precursor 
outburst had been observed \textcolor{black}{(see below)}, 
while precursors have been observed in 
almost all of ordinary SU UMa-type stars except for large-amplitude, 
rarely outbursting objects \textcolor{black}{\citep{war95book}}.  
In ordinary SU UMa-type 
stars, the disk radius exceeds the 3:1 resonance radius at the onset 
of their superoutbursts.  It takes one or two days for the tidal 
instability to grow up.  Hence, the cooling wave propagates in the disk 
until the tidal instability fully develops. In the meantime, the 
brightness gradually decreases like normal outbursts.  After the 
disk becomes eccentric, starting a superoutburst, the brightness 
rapidly increases \citep{osa02wzsgehump}.  However, in WZ Sge-type 
stars, the mass stored in quiescence is large and the cooling 
front does not propagate before appearance of superhumps 
\citep{osa03DNoutburst}.  This is \textcolor{black}{considered to be} the 
reason why the light curves on WZ Sge-type stars had shown no 
precursors.

   AL Com is one of the best known WZ Sge-type dwarf novae, 
and sometimes called ``a perfect twin'' of WZ Sge itself. The 
WZ Sge nature of the system in the modern sense 
was first elucidated by \citet{kat96alcom}, who detected early and 
ordinary superhumps having periods of $P_{\rm Esh} $ = 0.05666(2) d 
and $P_{\rm sh}$ = 0.05722(10) d, respectively, during the 1995 
superoutburst.  
The 1995 superoutburst was studied by many 
astronomers. \citet{how96alcom} performed $V$ and $I$ photometry 
and derived a positive $P_{\rm dot} ( = \dot{P} / P$; the time 
derivative of the superhump period), together with a hint of 
the presence of stage A and C. \citet{pat96alcom} 
pointed out that the mass of the secondary star might be below 
0.04 $M_{\solar}$. \citet{nog97alcom} studied superhump period 
changes using extensive data and obtained $P_{\rm dot} = +1.5(3.1) 
\times 10^{-5}$. AL Com also underwent superoutbursts in 2001 
May \citep{ish02wzsgeletter}, 2007 October \citep{uem08alcom}, 
and 2013 December \citep{Pdot6}. All of these superoutbursts 
showed early superhumps and rebrightenings.

   On 2015 March 4.582 UT, Kevin Hills reported an outburst of 
AL Com at a magnitude of $V=14.528$ (vsnet-alert 18377). 
Immediately after this report, the VSNET collaboration team 
\citep{VSNET} started a worldwide photometric campaign. 
Surprisingly, the present outburst turned out to be a superoutburst, 
despite the fact that the brightness was much fainter than 
those in past superoutbursts and that the previous superoutburst 
occurred only $\sim$450 d before. Here, we report 
our observation of AL Com during this unusual superoutburst.

\section{Observations and Results}

   Time-resolved CCD photometry was carried out by the VSNET 
collaboration team at seven sites. Table 1 (online only supplementary 
data) shows the log of photometric observations.
We also used the data downloaded from the AAVSO archive 
\footnote{$<{\rm http://www.aavso.org/data/download/}>$.}.
All of the observation times were converted to Barycentric 
Julian Date (BJD).
The magnitude scales of each site were adjusted to that of 
the Kolonica Saddle system (DPV in Table 1), where 
UCAC4 522-052504 (RA: 12h32m10.04s, Dec:+\timeform{14D20'15.4''}, 
$V$ = 13.51, \citep{ber64alcom}) was used as the comparison 
star. The constancy of the comparison star was checked by 
nearby stars in the same images.

\begin{figure}[htb]
\begin{center}
\FigureFile(80mm, 50mm){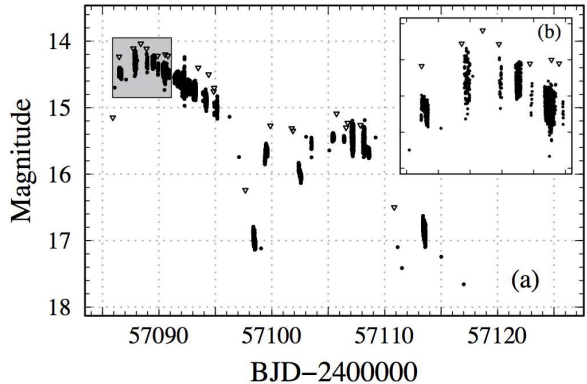}
\end{center}
\caption{(a) Overall light curve of AL Com (BJD $2457085 - 
2457120$). (b) Enlargement light curve around the precursor 
(BJD $2457086 - 2457091$). It is the enlarged view of the shaded 
box in panel (a). The black circles and inverted triangles 
represent CCD photometric observations and visual observations, 
respectively.}
\label{g-luminosity}
\end{figure}

   Figure \ref{g-luminosity}(a) represents the overall light 
curve of the 2015 outburst of AL Com.
As can be seen in Figure \ref{g-luminosity}(a), 
the duration of the plateau stage is ${\sim}$10 d. This 
duration is significantly shorter than those of previous 
superoutbursts, for which the plateau stages lasted for 20 d or 
more \citep{kat96alcom,ish02wzsgeletter,uem08alcom,Pdot6}. 
In addition, the maximum magnitude was 14.1 mag, about 2 mag 
fainter than those of previous superoutbursts. It also 
should be noted that the superoutburst was accompanied by a 
precursor (BJD 2457086; see Figure \ref{g-luminosity}(b)).  
The 2015 superoutburst of AL Com is the 
first case that a precursor was observed in WZ Sge-type dwarf novae. 
After the termination of the plateau stage, AL Com underwent 
a rebrightening. The present rebrightening was well classified 
as the type-A rebrightening (see also \cite{kat15wzsge}).

\begin{figure}[htb]
\begin{center}
\FigureFile(60mm, 70mm){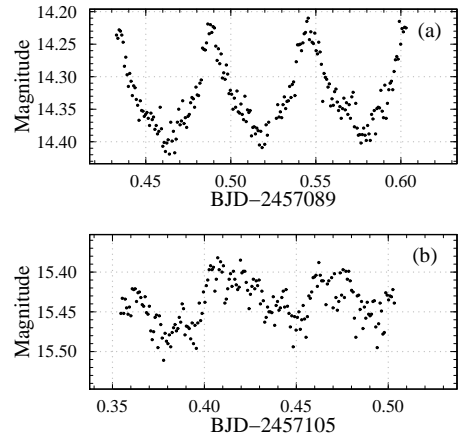}
\end{center}
\caption{Light curves of superhumps: (a) superhumps in the 
main outburst (BJD $2457089.40 - 2457089.65$) and (b) those in 
the rebrightening (BJD $2457105.35 - 2457105.50$).}
\label{s-luminosity}
\end{figure}

   Figure \ref{s-luminosity}(a) shows an 
enlarged light curve on BJD 2457089, on the first night 
when superhumps appeared. In combination with Figure 
\ref{g-luminosity}(a), the growth time scale of the superhumps 
can be determined as $\sim$3 d, an unprecedentedly short value 
among WZ Sge-type dwarf novae. Despite careful analyses before 
the appearance of the superhumps, no evidence for early 
superhumps was present. The present superoutburst of AL Com 
is also the first example in which early superhumps are absent 
in a superoutburst of WZ Sge-type dwarf novae.
In Figure \ref{s-luminosity}(b), an enlarged light curve on 
BJD 2457105 is depicted. We can see the regrowth of superhumps 
after the temporary disappearance 
of superhumps around the dip in the rebrightening phase in 
Figure \ref{s-luminosity}(b). 

\begin{figure}[htb]
\begin{center}
\FigureFile(70mm, 80mm){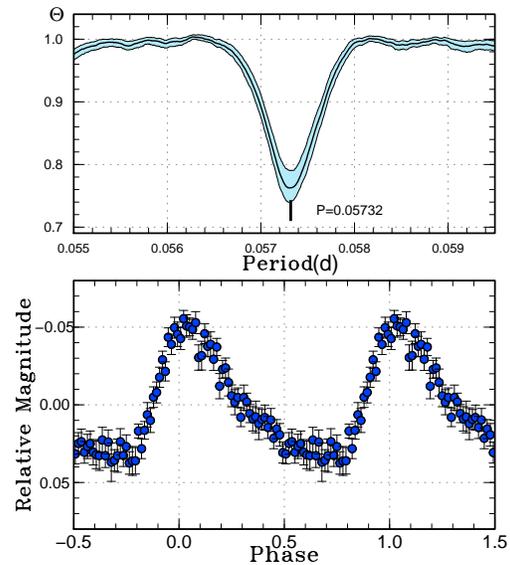}
\end{center}
\caption{Superhumps in the plateau stage of the 2015 outburst 
of AL Com (BJD $2457089.3 - 2457095.5$).  
The upper panel represents $\Theta$-diagram of our PDM 
analysis. The lower panel represents a phase-averaged 
profile.}
\label{s-pdm}
\end{figure}

   Next, we performed a period analysis by using the Phase 
Dispersion Minimization method (PDM: \cite{PDM}) during the 
plateau stage. In order to subtract a global trend from the 
light curve, we used locally weighted polynomical regression 
(LOWESS: \cite{LOWESS}). The 1$\sigma$ errors for the PDM 
analysis were calculated by the methods in \citet{fer89error} 
and \citet{Pdot2}. The resultant $\Theta$ diagram is shown in 
the upper panel of Figure \ref{s-pdm}. The obtained period of 
$P_{\rm sh}$ = 0.0573185(11) d well agrees with those of 
previous superoutbursts (\cite{nog97alcom,Pdot6}).

\begin{figure}[htb]
\begin{center}
\FigureFile(80mm, 100mm){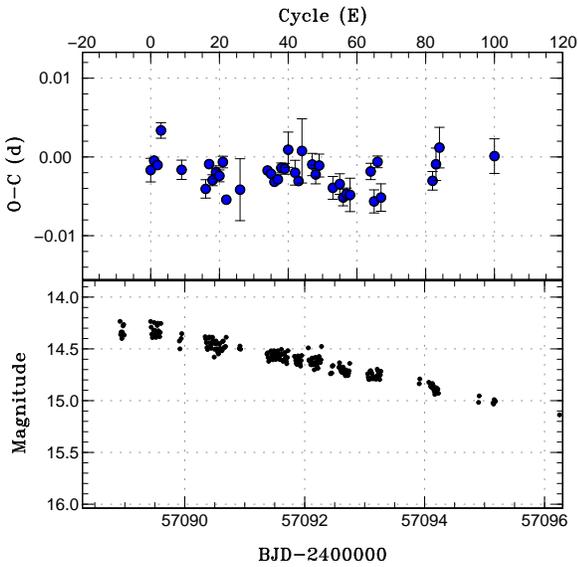}
\end{center}
\caption{The upper panel represents the $O - C$ curve 
of the superhump-maximum timings of AL Com (BJD $2457089.3 
- 2457095.5$). An emphemeris of BJD 2457089.432681$+$0.057318 
E was used for drawing this figure. The lower panel represents 
the light curve during BJD $2457089.3 - 2457095.5$. The 
horizontal axis in units of BJD and cycle number is common 
to both of upper and lower panels.}
\label{o-c}
\end{figure}

   We show the $O - C$ curve of superhump-maximum timings 
of AL Com in the upper panel of Figure \ref{o-c}. The light 
curve during BJD $2457089.3 - 2457095.5$ is shown in the 
lower panel of Figure \ref{o-c}. In the upper panel of 
Figure \ref{o-c}, the $O - C$ curve does not seem to be 
linear. Therefore, we regard superhumps in the 
plateau stage as stage B superhumps \citep{Pdot}. We could 
not find stage A and stage C superhumps in the upper panel 
of Figure \ref{o-c}. 
The time derivative of the superhump period $P_{\rm dot} 
= +1.5 (3.1) \times 10^{-5}{\rm s}~{\rm s}^{-1}$ was 
recorded in stage B. This value agrees with those obtained 
in previous superoutbursts (\cite{nog97alcom,Pdot}).

\section{Discussion}

\subsection{Superoutburst}

   One of the most surprising results is that AL Com exhibited 
a superoutburst with a recurrence time of ${\sim}$450 d, 
despite the fact that the AL Com is classified as WZ Sge-type 
dwarf novae and that a supercycle of the system has been 
${\sim}$6 yr over the past 20 years. WZ Sge 
stars having a short supercycle include AL Com and EZ Lyn. 
EZ Lyn has the shortest recorded interval between superoutbursts 
of ${\sim}$4 yr, based on the reported superoutbursts 
(\cite{pav12ezlyn}; \cite{iso15ezlyn}). The 
present superoutburst of AL Com has established the shortest 
interval of superoutbursts among WZ Sge-type dwarf novae. 
At present, the reason why AL Com underwent a superoutburst 
with such a short interval is unclear; it was likely disk 
instability was somehow triggered earlier than in past 
superoutbursts.

\subsection{Precursor and Early Superhumps}

   It should be noted that a precursor was observed for the 
first time in WZ Sge-type dwarf novae (see Figure 
\ref{g-luminosity}(b)).  \textcolor{black}{In spite of good coverlage 
around the onset of outbursts,} the absence of a precursor in 
WZ Sge-type dwarf novae is confirmed in e.g., GW Lib \citep{hir09gwlib}, 
V455 And \citep{mat09v455and}, and WZ Sge itself \citep{ish02wzsgeletter}. 
These systems show large-amplitude and long-lasting superoutbursts 
compared with the present superoutburst of AL Com. 
According to the thermal-tidal instability 
model, the disk radius of the main superoutburst in which 
precursors are observed critically exceeds the 3:1 resonance 
radius during the precursor outburst (see, figure 4 of 
\cite{osa03DNoutburst}). In combination with the thermal-tidal 
instability model and the past investigation of the disk radius 
variation as reviewed in section 1, the observed precursor 
in the present superoutburst suggests that the maximum radius 
of the accretion disk is slightly larger than the 3:1 resonance 
radius.

   It is also noteworthy that the present superoutburst lacked 
early superhumps. In other words, superhumps appeared $\sim$3 d 
from the onset of the superoutburst. The appearance of the 
superhumps in the present superoutburst was about a week 
earlier than those in previous superoutbursts. 
In fact, \citet{how96alcom} reported that 
it took 10 d for superhumps to develop during the 1995 
superoutburst. According to \citet{lub91SHa}, the growth 
time of the tidal instability is inversely proportional to 
the square of the mass ratio. \citet{osa95wzsge} also 
followed this interpretation. Instead, 
\citet{osa02wzsgehump} and \citet{osa03DNoutburst} 
proposed that early superhumps are manifestation of the 
2:1 resonance radius and that the 2:1 resonance suppresses the 
development of superhumps. The observed 
precursor and the lack of early superhumps can be explained as the 
result of condition that the maximum radius is below the 2:1 
resonance. 
The striking difference of the development of ordinary superhumps 
on the same object (AL Com) indicates that the mass ratio cannot 
be the main cause of the long delay of appearance of ordinary 
superhumps in WZ Sge-type dwarf novae.  We can interpret that the 2:1 
resonance (and its manifestation, early superhumps) suppresses 
the development of ordinary superhumps.

\subsection{Superhump Period Changes}

   From the $O - C$ diagram, we derived $P_{\rm dot} = 
+1.5 (3.1) \times 10^{-5}{\rm s}~{\rm s}^{-1}$. This 
value is in good agreement with those of previous 
works (e.g., \cite{Pdot}). Based on the extensive data 
of $P_{\rm dot}$, \citet{Pdot} suggested that each object 
has the same $P_{\rm dot}$ regardless of whether or not 
the superoutburst is accompanied by a precursor in ordinary 
SU UMa-type dwarf novae. The present superoutburst further 
supports the result of \citet{Pdot} even in WZ Sge-type ones.

\subsection{Rebrightening}

   Although the plateau stage of the present superoutburst 
was similar to those of SU UMa-type dwarf novae, a long 
rebrightening (type-A rebrightening) was observed in the 
present outburst of AL Com.  
AL Com showed the type-A (or type A/B) 
rebrightening also in all of the previous superoutbursts 
(\cite{nog97alcom}; \cite{ish02wzsgeletter}; \cite{uem08alcom}; 
\cite{Pdot6}; \cite{kat15wzsge}).  
In the 2015 outburst, 
the rebrightening type was the same as previous 
ones in spite that the stored disk mass just before the 
outburst should have been much less than previous ones, 
which means that the rebrightening type of AL Com is the 
same regardless of the extent of outburst.  
It may be possible that 
the matter remained outside the 3:1 resonance if the radius 
of the 3:1 resonance is well inside the Roche lobe.  The 
uniqueness of the type of the rebrightening within the same 
object is recorded not only in AL Com but also in other WZ 
Sge-type dwarf novae.  For example, EZ Lyn showed only 
type-B rebrightening. Another example is the 2003-04 and 
2013 superoutbursts of UZ Boo, which showed only type-B 
rebrightening \citep{kat15wzsge}.  
\citet{Pdot6} studied the 
relation between $P_{\rm dot}$ and $P_{\rm orb}$ in terms of 
the type of the rebrightening, in which each type of the 
rebrightening pattern is clustered on the $P_{\rm dot} - 
P_{\rm orb}$ diagram (see figure 83 of \citet{Pdot6}).  In 
conjunction with the present study and the $P_{\rm dot} - 
P_{\rm orb}$ diagram, the type of the rebrightening seems to 
be inherent in the same system.  This 
reproducibility of rebrightening type needs to be explained 
by theoretical investigations. 

   As shown in Figure \ref{s-luminosity}(b), 
a regrowth of the superhumps was observed during the 
rebrightenings. This suggests that the accretion disk again 
extended beyond the 3:1 resonance radius over the course of 
the rebrightenings.

\section{Summary}

We have reported unusual superoutburst and 
rebrightening of AL Com in 2015 March. We summarize the most 
important results of our observations as follows: 
\begin{enumerate}
\item AL Com showed a superoutburst with an interval of 
${\sim}$450 d. This interval is the shortest record among 
WZ Sge-type dwarf novae. 
\item
The light curve showed a precursor and lacked early superhumps, 
which is the first example in the superoutbursts of 
WZ Sge-type dwarf novae. The presence of the precursor and the 
lack of early superhumps suggest that the maximum disk radius 
of this superoutburst barely exceeded the 3:1 resonance and did 
not reach the 2:1 resonance radius following the theory proposed 
by \citet{osa02wzsgehump} and \citet{osa03DNoutburst}.
\item 
AL Com showed the type-A rebrightening in the present 
superoutburst, despite the fact that the main superoutburst 
mimicked the typical one in ordinary SU UMa-type dwarf novae, 
and despite that the mass stored in the accretion disk at the 
onset of the superoutburst is supposed to be much smaller than 
those in previous superoutbursts of AL Com.
\end{enumerate}

\vskip 5mm

This work was supported by a Grant-in-Aid ``Initiative for 
High-Dimensional Data-Driven Science through Deepening of 
Sparse Modeling'' from the Ministry of Education, Culture, 
Sports, Science and Technology (MEXT) of Japan (25120007).
We are thankful to many amateur observers for providing a 
lot of data used in this research.
We acknowledge with thanks the variable-star observations 
from the AAVSO International Database contributed by 
observers worldwide and used in this research.
We are grateful to the anonymous referee for insightful 
comments.

\newcommand{\noop}[1]{}

\begin{table*}[b]
\caption{Log of observations of AL Com in 2015}
\label{log}
\begin{center}
\begin{tabular}{rrrrrccr}
\hline
${\rm Start}^{*}$ & ${\rm End}^{*}$ & ${\rm Mag}^{\dagger}$ & ${\rm Error}^{\ddagger}$ & $N^{\S}$ & ${\rm Obs}^{\parallel}$ & ${\rm Band}^{\#}$ & exp[s] \\ \hline
86.4595 & 86.6849 & 14.253 & 0.002 & 307 & RPc & $V$ & 60 \\
87.8162 & 87.9366 & 14.116 & 0.011 & 72 & AAVSO & $V$ & 120 \\
89.4328 & 89.6030 & 14.326 & 0.004 & 214 & DPV & $C$ & 60 \\
90.3395 & 90.4716 & 14.437 & 0.004 & 170 & DPV & $C$ & 60 \\
90.4260 & 90.6889 & 14.261 & 0.003 & 285 & RPc & $V$ & 60 \\
90.4903 & 90.5754 & 14.225 & 0.006 & 120 & AAVSO & $V$ & 60 \\
91.3727 & 91.6040 & 14.572 & 0.002 & 285 & DPV & $C$ & 60 \\
91.5368 & 91.7189 & 14.356 & 0.003 & 248 & RPc & $V$ & 60 \\
91.8336 & 91.9516 & 14.418 & 0.004 & 100 & AAVSO & $V$ & 120 \\
91.8388 & 91.9311 & 14.376 & 0.004 & 92 & DKS & $C$ & 60 \\
92.0640 & 92.2760 & 14.215 & 0.004 & 541 & Kis & $I{\rm c}$ & 30 \\
92.4319 & 92.7412 & 14.498 & 0.003 & 267 & RPc & $V$ & 60 \\
92.7250 & 92.7544 & 14.454 & 0.009 & 29 & DKS & $C$ & 60 \\
93.0458 & 93.2710 & 14.588 & 0.002 & 511 & Kis & $C$ & 30 \\
94.0966 & 94.2286 & 14.718 & 0.003 & 347 & Kis & $C$ & 30 \\
95.1349 & 95.1804 & 14.844 & 0.005 & 121 & Kis & $C$ & 30 \\
98.3414 & 98.5572 & 16.985 & 0.006 & 140 & DPV & $C$ & 60 \\
99.3920 & 99.6271 & 15.643 & 0.003 & 302 & DPV & $C$ & 60 \\
102.3807 & 102.6172 & 15.993 & 0.005 & 159 & DPV & $C$ & 60 \\
103.5077 & 103.5213 & 15.927 & 0.015 & 20 & AAVSO & $V$ & 60 \\
105.3546 & 105.5033 & 15.439 & 0.002 & 191 & DPV & $C$ & 60 \\
106.3781 & 106.4277 & 15.466 & 0.003 & 65 & DPV & $C$ & 60 \\
107.0658 & 107.2406 & 15.276 & 0.004 & 457 & Kis & $C$ & 30 \\
108.0868 & 108.2301 & 15.326 & 0.006 & 247 & Kis & $C$ & 30 \\
108.2852 & 108.5968 & 14.823 & 0.002 & 273 & CRI & $C$ & 90 \\
108.4375 & 108.6461 & 15.331 & 0.002 & 290 & RPc & $C$ & 60 \\
108.3752 & 108.5311 & 15.313 & 0.002 & 194 & IMi & $C$ & 60 \\
113.3872 & 113.6240 & 16.518 & 0.006 & 257 & IMi & $C$ & 60 \\
\hline
\multicolumn{8}{l}{$^{*}{\rm BJD}-2457000.0$.}\\
\multicolumn{8}{l}{$^{\dagger}$Mean magnitude.}\\
\multicolumn{8}{l}{$^{\ddagger}1\sigma$ of mean magnitude.}\\
\multicolumn{8}{l}{$^{\S}$Number of observations.}\\
\multicolumn{8}{l}{$^{\parallel}$Observer's code: RPc (Roger D. Pickard),}\\
\multicolumn{8}{l}{AAVSO (AAVSO observers: Boardman James and Rodda Anthony),}\\
\multicolumn{8}{l}{DPV (Pavol A. Dubovsky), DKS (Shawn Dvorak), Kis (Seiichiro Kiyota),}\\
\multicolumn{8}{l}{CRI (Elena P. Pavlenko and Aleksei A. Sosonovskij), and IMi (Ian Miller)}\\
\multicolumn{8}{l}{$^{\#}$Filter. ``$V$'' means $V$ filter,``$C$'' means no filter (clear),}\\
\multicolumn{8}{l}{and ``$I{\rm c}$'' means $I$ filter with unfilterd zeropoint.}\\
\multicolumn{8}{l}{}
\end{tabular}
\end{center}
\end{table*}

\end{document}